\documentclass[showpacs,preprintnumbers,superscriptaddress,amsmath,amssymb]{revtex4-2}

\usepackage[colorlinks=true,bookmarks=false,citecolor=red,urlcolor=blue,linkcolor=blue]{hyperref}
\usepackage[top=2.5cm, bottom=2.5cm,left=2.5cm, right=2.5cm]{geometry}
\usepackage{amsmath,amssymb}
\usepackage{graphicx}
\usepackage{verbatim}
\usepackage{enumitem}
\usepackage{color}
\usepackage[english]{babel}
\usepackage{bm}
\usepackage{natbib}
\usepackage[symbol]{footmisc}
\usepackage{textcomp}
\usepackage{epstopdf}
\usepackage{setspace}

\newcommand{\RNum}[1]{\uppercase\expandafter{\romannumeral #1\relax}}
\def\P{\mathcal{P}}

\def\br{{\bf r}}

\def\siamjam{SIAM J. Appl. Math.}
\def\siamjma{SIAM J. Math. Anal.}

\def\jcp{J. Comput. Phys.}

\def\opex{Opt. Express}
\def\ol{Opt. Lett.}
\def\josaa{J. Opt. Soc. Am. A}
\def\josab{J. Opt. Soc. Am. B}

\def\prb{Phys. Rev. B}
\def\pre{Phys. Rev. E}
\def\prl{Phys. Rev. Lett.}

\begin{document}

\title{An efficient method for calculating resonant modes in biperiodic photonic structures}

\author{Nan Zhang}
\thanks{Corresponding Author}
\email{nzhang234-c@my.cityu.edu.hk} 
\affiliation{Department of Mathematics, City University of Hong Kong, Hong Kong}

\author{Ya Yan Lu}
\affiliation{Department of Mathematics, City University of Hong Kong, Hong Kong}
\date{\today}

\begin{abstract}
Many photonic devices, such as photonic crystal slabs, cross gratings, and periodic metasurfaces,
are biperiodic structures with two independent periodic directions, and are sandwiched between two homogeneous media.
Many applications of these devices are closely related to resonance phenomena.
Therefore, efficient computation of resonant modes is crucial in device design and structure analysis.
Since resonant modes satisfy outgoing radiation conditions,
perfectly matched layers (PMLs) are usually used to truncate the unbounded spatial variable perpendicular to the periodic directions.
In this paper, we develop an efficient method without using PMLs to calculate resonant modes in biperiodic structures.
We reduce the original eigenvalue problem to a small matrix nonlinear eigenvalue problem which is solved by the contour integral method.
Numerical examples show that our method is efficient with respect to memory usage and CPU time, free of spurious solutions,
and determines degenerate resonant modes without any difficulty. 
\end{abstract}

\maketitle

\section{Introduction}
In recent years, biperiodic photonics structures have attracted much attention due to their easy fabrication, tailorability, and wide
range of functionalities~\cite{Nedelec1991,Dobson1994,Joannopoulos2000,Johnson1999,Fan2002,Joannopoulos2008,Sakoda2005,Hsu2013,Zhen2015}. 
A biperiodic structure is a three-dimensional (3D) structure that is periodic in two independent directions in the $xy$ plane, bounded in $z$, and sandwiched between two homogeneous media~\cite{Nedelec1991,Dobson1994}.
Common examples include a triangular lattice of dielectric spheres~\cite{Bulgakov2019} and a dielectric slab with a square lattice of air holes~\cite{Joannopoulos2008,Hsu2013}. 
These structures also become a versatile platform for studying novel wave phenomena such as bound states in the continuum~\cite{Hsu2013,bonnet1994,Hsu16NRM} and exceptional points~\cite{Zhen2015,kato1966}.
Some biperiodic structures may support guided modes which are eigensolutions of Maxwell's equations without any source and decays exponentially as $|z|\rightarrow\infty$~\cite{Johnson1999,Joannopoulos2008}.
In lossless structures, a guided mode usually has a real Bloch wavevector and a real angular frequency $\omega$. 
They can form bands and typically exist below the lightline~\cite{Johnson1999,Joannopoulos2008,Hsu2013}.
Above the lightline there exist resonant modes which are divergent in $z$~\cite{Fan2002}.
A resonant mode is an eigensolution of Maxwell's equations satisfying an outgoing radiation condition~\cite{richtmyer1939,Shenk1971,Shenk1972,Lenoir1992,Shipman2003,Alu2019,Shipman2005,Lee2012}. 
It has a real Bloch wavevector and a complex frequency $\omega$ due to radiation loss~\cite{Fan2002,Shipman2003}.
A resonant mode is equipped with a quality factor $Q$ which indicates the ratio of the stored energy to the radiation loss~\cite{richtmyer1939,Haus1984}.
A high-$Q$ resonance can be used to enhance nonlinear optical effects~\cite{Carletti18PRL}, design low-threshold lasers~\cite{Kodigala17Nature}, 
and develop sensors with high sensitive and large figure of merit~\cite{kim2015}, etc.
Efficient computation of resonant modes is essential for analyzing photonic structures and designing photonic devices~\cite{itoh1977,wei1988,hwang1998,Jay2008,zou2009,Lin2019,Carlos2020}.

Since resonant modes in a biperiodic structure are quasi-periodic, they can be solved in a unit cell that is unbounded in $z$~\cite{Fan2002}.
A variety of numerical methods have been developed for computing resonant modes.
It is possible to find resonant modes by solving time-domain Maxwell's equations~\cite{Taflove2000,Johnson2010MEEP}.
More specifically, by recording the electromagnetic fields at certain locations, 
the frequencies of resonant modes can be extracted using signal processing tools.
The electromagnetic fields of the resonant modes can also be obtained in the time-domain simulation 
using a narrow-bandwidth (long-time) pulse as a source term.
On the other hand, the resonant modes, by definition, are solutions of eigenvalue problems, 
and thus, they are more naturally solved from the frequency-domain Maxwell's equations.
A popular approach is to discretize the Maxwell's equations by a numerical method, such as the finite element (FEM), 
finite difference, or spectral method,
and truncate the $z$ variable by perfectly matched layers (PMLs)~\cite{Carlos2020,Kim2009,Parisi2012,Shin2012,Johnson2001,Shi2004,Burger2020,Binkowski2024,Nannen2018,Lalanne2023}.
It always leads to a large generalized matrix eigenvalue problem which can be handled by iterative eigenvalue solvers such as the Arnoldi method~\cite{Arnoldi1951,Moler1973,Golub2013}.  
However, it is not a straightforward matter to determine a proper setting of PML parameters,
and it is costly to detect spurious solutions resulted from domain discretization and PML truncation~\cite{Kim2009,Engstrom2021,Lalanne2023}.
Some methods introduce artificial boundaries with non-local boundary conditions to truncate the unit cell to a bounded domain without using PMLs~\cite{Lalanne2023,Tausch2000,Fliss2013,Araujo2018,Jiang2022}. 
Compared with those methods based on PMLs, the truncated domain is much smaller. 
The differential equations in the bounded domain can be discretized by standard numerical methods.
These methods give rise to large matrix nonlinear eigenvalue problems (NEPs) 
which can be solved by contour integral method~\cite{Asakura2009,Beyn2012,Guttel2017} or Newton-type methods~\cite{Guttel2017,Kublanovskaya1969,Kublanovskaya1970,Kressner2009}.

There are also some methods utilizing the particular natures of structures to reduce the workload.
For piecewise homogeneous structures, the boundary integral equation method can be used to solve the resonant modes~\cite{Glisson1983,Tausch2013,lin2023}. 
The method reduces the original 3D problem to a 2D problem on the boundary or the interfaces of the given structure. 
Since Green's function depends on the frequency, the boundary integral equation method gives rise to matrix NEPs. 
In addition, for some multilayer structures where the Maxwell's equations is separable in different regions, the resonant modes can be calculated by using the mode matching method~\cite{Tikhodeev2002,Liu2012,Shi2016}.
Since the mode matching method must calculate the eigenmodes in each separable region and these eigenmodes depend on the frequency,
the method also gives rise to matrix NEPs.
For the boundary integral equation method and the mode matching method, the size of the resulting matrix is much smaller than that in methods based on domain discretization.

In this paper, we develop a new method for calculating the resonant modes in arbitrary biperiodic structures.
The unit cell is divided into a bounded interior and two unbounded exterior subdomains, and we use
the transverse impedance (TI) operators in each subdomains to build a NEP.
Since for any given Bloch wavevector, the resonant mode is quasi-periodic and smooth at the interfaces between the subdomains,
the TI operators can be approximated by relatively small matrices.
Consequently, the original eigenvalue problem of the Maxwell's equation is reduced to a small matrix NEP which can be solved efficiently by the contour integral method.
The exterior TI matrix is obtained directly from the Rayleigh expansion of fields in the exterior subdomains.
The interior TI matrix is computed using a FEM. 
Moreover, we show that the interior TI matrix is exactly the Schur complement of a large argument matrix,
and it can be calculated by a fast and memory-efficient direct solver. 
Our numerical examples indicate that even when the argument matrix is very large, a relatively small TI matrix is sufficient to retain the accuracy.
We also show that this method is efficient with respect to memory usage and CPU time, free of spurious solutions, and determines degenerate resonant modes without any difficulty.

The paper is organized as follows: In section 2, we introduce some background and define resonant modes in biperiodic structures.
In section 3, we convert the original Maxwell eigenvalue problem to an operator NEP.
In sections 4 and 5, we approximate the TI operators by matrices and calculate the resonant modes using the contour integral method.
Numerical examples are provided in section 6 to demonstrate the efficiency of our method.
\section{Resonant modes in biperiodic structures}
We consider a three-dimensional (3D) non-magnetic structure that is bounded in $z$ by $|z|<d$ for some $d>0$, sandwiched between two homogeneous media, 
and periodic in $x$ and $y$ with period $L$. 
The dielectric function $\varepsilon(x,y,z)$ of the structure and the surrounding media satisfies
\begin{equation}
	\varepsilon({\bf r})=\varepsilon \left(x+L,y,z\right) = \varepsilon (x,y+L,z),
\end{equation}
and $\varepsilon({\bf r})=\varepsilon^\pm$ if $\pm z>d$, where ${\bf r}=(x,y,z)$ and $\varepsilon^\pm$ are constants satisfying $\varepsilon^\pm\geq 1$.
We study time-harmonic electromagnetic waves that depend on time $t$ as $\exp(-{\rm i}\omega t)$, where $\omega$
is the angular frequency.
The Maxwell's equations in frequency-domain can be written as
\begin{equation}
	\nabla\times{\bf E}={\rm i}\frac{\omega}{c}{\bf H},\quad \nabla\times{\bf H}=-{\rm i}\frac{\omega}{c}\varepsilon({\bf r}){\bf E},
\end{equation}
where $c$ is the speed of light in vacuum, ${\bf E}({\bf r})$ is the electric field, and ${\bf H}({\bf r})$  is a scaled magnetic field (magnetic field multiplied by free-space impedance).

In the biperiodic structure, the electric field $\bf E$ of a Bloch mode with Bloch wavevector $[\alpha;\beta;0]$ satisfies the following quasi-periodic condition
\[
{\bf E}(x+mL,y+nL,z)=e^{{\rm i}(\alpha m+\beta n)L}{\bf E}({\bf r}),
\]
where $m$ and $n$ are arbitrary integers.
In a unit cell $\Omega=S\times\mathbb{R}$, where $S=(-L/2,L/2)\times(-L/2,L/2)$, the Bloch mode corresponds to an eigenpair $(\omega,{\bf E})$ which satisfies the equation
\begin{equation}\label{WaveEqu}
	\nabla\times\nabla\times{\bf E}-\left(\frac{\omega}{c}\right)^2\varepsilon({\bf r}){\bf E}=0,\quad{\bf r}\in\Omega,
\end{equation}
the quasi-periodic boundary condition
\begin{equation}\label{QPBC}
	{\bf E}(L/2,y,z)=e^{{\rm i}\alpha L}{\bf E}(-L/2,y,z),\quad{\bf E}(x,L/2,z)=e^{{\rm i}\beta L}{\bf E}(x,-L/2,z),
\end{equation}
and a proper boundary condition as $z\rightarrow\pm\infty$.
A resonant Bloch mode satisfies an outgoing radiation condition in $z$.
It has a real Bloch wavevector and a complex frequency $\omega$ with $\mbox{Im}(\omega)<0$.
The quality factor ($Q$-factor)  of the resonant mode is defined by $Q=-0.5\mbox{Re}(\omega)/\mbox{Im}(\omega)$.
In the regions $|z|>d$, an electric field ${\bf E}$ of the resonant mode has the following Rayleigh expansion
\begin{equation}\label{EXP}
	{\bf E}(\br)=\sum_{m,n=-\infty}^{+\infty}{\bf E}_{mn}^{\pm}F_{mn}(x,y)e^{\pm {\rm i}\gamma^\pm_{mn}z},\quad\pm z>d,
\end{equation}
where ${\bf E}_{mn}^{\pm}$ are the Fourier coefficients corresponding to the Fourier terms $F_{mn}(x,y)=\exp\left[{{\rm i}(\alpha_mx+\beta_ny)}\right]$, $\alpha_m=\alpha+2\pi m/L$, $\beta_n=\beta+2\pi n/L$,  $\gamma_{mn}^\pm=\sqrt{(\omega/c)^2\varepsilon^\pm-\alpha_m^2-\beta_n^2}$, 
and the complex square root follows a branch cut along the negative imaginary axis. 
The outgoing radiation condition requires that at least one of $\gamma_{mn}^\pm$ has a positive real part and a negative imaginary part.
Thus, the resonant mode is unbounded in $\Omega$.
If the structure is symmetric in $z$, i.e., $\varepsilon(x,y,z)=\varepsilon(x,y,-z)$,
we can find electric fields whose components are either even or odd in $z$.
More precisely, we introduce a parity operator $\P$ such that
\begin{equation}
	\P{\bm w}=\left[w_x(x,y,-z);w_y(x,y,-z);-w_z(x,y,-z)\right]
\end{equation}
for any vector function ${\bm w}({\bf r})$.
If the resonant mode is non-degenerate, i.e., there is only one linearly independent electric field, we have either $\P{\bf E}={\bf E}$ ($\P{\bf E}_{mn}^+={\bf E}_{mn}^-$) or $\P{\bf E}=-{\bf E}$ ($\P{\bf E}_{mn}^+=-{\bf E}_{mn}^-$).
For a doubly-degenerate resonant mode, we can find two linearly independent fields, ${\bf E}_1$ and ${\bf E}_2$, such that $\P{\bf E}_j={\bf E}_j$ or $\P{\bf E}_j=-{\bf E}_j$, $j=1,2$.

The real part of the complex frequency $\omega$ of a resonant mode typically satisfies 
\begin{equation}
	c\sqrt{\alpha^2+\beta^2}<\max\left\{\mbox{Re}(\omega)\sqrt{\varepsilon_-},\mbox{Re}(\omega)\sqrt{\varepsilon_+}\right\}.
\end{equation}
Importantly, in the region given by above inequality, there may exist special resonant modes with $\mbox{Im}(\omega)=0$, and they are so-called bound states in the continuum (BICs).
Notice that a BIC has a real frequency and does not radiate power to infinity.
Therefore, the coefficients ${\bf E}_{mn}^\pm$ of a BIC corresponding to the radiation terms with $\mbox{Re}(\gamma_{mn}^\pm)>0$ must be zero.
\section{Exterior and interior transverse impedance operators}
In this section, we introduce transverse impedance (TI) operators which map transverse magnetic fields to transverse electric fields at planes $\Gamma_{\pm}=S\times\{z=\pm D\}$ for some $D>d$.
These operators will be used to convert the eigenvalue problem (\ref{WaveEqu}) and (\ref{QPBC}) to an operator NEP.
The planes $\Gamma_\pm$ divide the unit cell $\Omega$ into three subdomains $\Omega_+$, $\Omega_-$ and $\Omega_D$, given by
\begin{equation}
	\Omega_+=S\times(D,+\infty),\quad\Omega_-=S\times(-\infty,-D),\quad\Omega_D=S\times(-D,D).
\end{equation}
For any tangential magnetic fields given on $\Gamma_{\pm}$ and satisfying Eq. (\ref{QPBC}),
we have the Fourier expansion
\begin{equation}\label{FourierH}
	{\bf H}_{\tau}^\pm=\sum_{m,n=-\infty}^{\infty}{\bf H}_{\tau,mn}^\pm F_{mn}(x,y),
\end{equation}
where ${\bf H}_{\tau}^\pm=\pm[- H_y^\pm;H_x^\pm;0]$ and ${\bf H}_{\tau,mn}^\pm$ are the Fourier coefficients.
Corresponding to the electromagnetic fields satisfying Eqs.~(\ref{QPBC}) and (\ref{EXP}) in $\Omega_+$ or $\Omega_-$,  we define the exterior TI operators $\mathcal{G}_{\rm ext}^\pm$ by 
\begin{equation}\label{FourierE}
	\mathcal{G}_{\rm ext}^\pm{\bf H}_{\tau}^\pm={\bf E}_{\tau}^\pm=\sum_{m,n=-\infty}^{\infty}{\bf E}_{\tau,mn}^\pm F_{mn}(x,y),
\end{equation}
where  ${\bf E}_{\tau}^\pm=[E_x^\pm;E_y^\pm;0]$, ${\bf E}_{\tau,mn}^\pm$ are the Fourier coefficients, and
\begin{equation}
	\left[\begin{aligned}
		E^\pm_{x,mn}\\
		E^\pm_{y,mn}
	\end{aligned}\right]=\pm{\bf T}^\pm_{mn}\left[\begin{aligned}
		-H^\pm_{y,mn}\\
		\;H^\pm_{x,mn}
	\end{aligned}\right],\quad 	{\bf T}^\pm_{mn}=-\frac{{\rm i}c}{\omega\varepsilon^\pm{\gamma}^\pm_{mn}}\begin{bmatrix}
		(\omega/c)^2-\alpha_m^2 & -\alpha_m\beta_n\\
		-\alpha_m\beta_n & (\omega/c)^2-\beta_n^2
	\end{bmatrix}.
\end{equation}

For electromagnetic fields satisfying Eq.~(\ref{QPBC}) in $\Omega_D$,
we define the interior TI operator $\mathcal{G}_{\rm int}$ by 
\begin{equation}\label{intFourier}
	\mathcal{G}_{\rm int}{\bm h}={\bm e}=\sum_{m,n=-\infty}^{\infty}{\bm e}_{mn}F_{mn}(x,y),
\end{equation}
where 
\[
{\bm h}=\begin{bmatrix}
	{\bf H}_{\tau}^+\\
	{\bf H}_{\tau}^-
\end{bmatrix},\quad{\bm e}=\begin{bmatrix}
	{\bf E}_{\tau}^+\\
	{\bf E}_{\tau}^-
\end{bmatrix},\quad{\bm e}_{mn}=\begin{bmatrix}
	{\bf E}_{\tau,mn}^+\\
	{\bf E}_{\tau,mn}^-
\end{bmatrix}.
\]
We emphasize that the exterior TI operators $\mathcal{G}_{\rm ext}^\pm$ have explicit expressions since the fields in $\Omega_\pm$ have Rayleigh expansions.
On the other hand, for a generic structure, the fields in $\Omega_D$ do not have an explicit expression. Thus, the interior TI operator $\mathcal{G}_{\rm int}$ has to be calculated by a numerical method.

Using the interior and exterior TI operators,  we can convert the eigenvalue problem for resonant modes to the following NEP
\begin{equation}\label{nonlineareig}
	\mathcal{G}(\omega){\bm h}=0,
\end{equation}
where 
\[
\mathcal{G}=\mathcal{G}_{\rm int}-\begin{bmatrix}
	\mathcal{G}^+_{\rm ext}& \\
	& \mathcal{G}^-_{\rm ext}
\end{bmatrix}.
\]

An eigenpair $(\omega,{\bm h})$ of this NEP corresponds to the frequency and tangential magnetic fields of a resonant mode.
The electromagnetic fields of the resonant mode in the unit cell can be obtained from the eigenfunction $\bm h$.
In $\Omega_D$, the fields can be obtained by solving the Maxwell's equations with quasi-periodic boundary condition (\ref{QPBC}) and the given transverse magnetic fields.
In the exterior domains $\Omega_\pm$, the fields can be obtained from the Rayleigh expansions using definition of exterior TI operators.
\section{Nonlinear matrix eigenvalue problem}
In this section, we use matrices to approximate the operators $\mathcal{G}^\pm_{\rm ext}$ and $\mathcal{G}_{\rm int}$ by truncating
the infinite series in Eqs.~(\ref{FourierH}), (\ref{FourierE}) and  (\ref{intFourier}) to finite sums, and reduce the NEP~(\ref{nonlineareig}) to a matrix NEP.
For quasi-periodic tangential magnetic fields ${\bf H}_{\tau}^\pm$ approximated by
\begin{equation}\label{Approximate}
	{\bf H}_{\tau}^\pm\approx \sum_{m,n=-M}^{M}{\bf H}_{\tau,mn}^\pm F_{mn}(x,y),
\end{equation}
where $M$ is a positive integer, we can approximate ${\mathcal{G}}_{\rm ext}^\pm{\bf H}_{\tau}^\pm$ and ${\mathcal{G}}_{\rm int}{\bm h}$ by
\begin{equation}\label{Truextint}
	{\mathcal{G}}_{\rm ext}^\pm{\bf H}_{\tau}^\pm\approx \sum_{m,n=-M}^{M}{\bf E}_{\tau,mn}^\pm F_{mn}(x,y),\quad
	\mathcal{G}_{\rm int}{\bm h}\approx \sum_{m,n=-M}^{M}{\bm e}_{mn}F_{mn}(x,y).
\end{equation}
The above formulas can be written in compact forms and the operator NEP (\ref{nonlineareig}) is reduced to a matrix NEP:
\begin{equation}\label{MatrixNEP}
	{\bf T}(\omega){\bf h}=0,\;{\bf T}={\bf T}_{\rm int}-{\bf T}_{\rm ext},
\end{equation}
where ${\bf T}_{\rm ext}$ and ${\bf T}_{\rm int}$ are ${P}\times {P}$ exterior and interior TI matrices, respectively, ${P}=4(2M+1)^2$, and $\bf h$ is a ${P}\times 1$ vector which is a collection of the Fourier coefficients of tangential magnetic fields.
The vectors $\bf h$ and the matrix ${\bf T}_{\rm ext}$ are given by
\begin{equation}\label{vectorh}
	{\bf h}=\begin{bmatrix}
		- { H}^+_{y,-M,-M}\\
		\;\; { H}^+_{x,-M,-M}\\
		\vdots\\
		\;\;{ H}^-_{y,-M,-M}\\
		-{ H}^-_{x,-M,-M}\\
		\vdots\\
		\;\;{ H}^-_{y,M,M}\\
		- { H}^-_{x,M,M}\\
	\end{bmatrix}
\end{equation}
and
\begin{equation}\label{matrixT}
	{\bf T}_{\rm ext}=\begin{bmatrix}
		{\bf T}_{\rm ext}^+& \\
		& {\bf T}_{\rm ext}^-
	\end{bmatrix},\quad
	{\bf T}_{\rm ext}^\pm=\begin{bmatrix}
		{\bf T}_{-M,-M}^\pm& & & &\\
		& {\bf T}_{-M,-M+1}^\pm & & &\\
		& & \ddots & &\\
		& & & & {\bf T}_{M,M}^\pm
	\end{bmatrix}.
\end{equation}

Next, we describe the steps for computing the matrix ${\bf T}_{\rm int}$.
Let ${\bf h}_1,\cdots,{\bf h}_{{P}}$ be the columns of the ${P}\times {P}$ identity matrix.
For each unit vector ${\bf h}_p$, $1\leq p\leq P$, according to Eqs.~(\ref{Approximate}) and (\ref{vectorh}),
we have tangential magnetic fields ${\bm h}_p=[{\bf H}_{\tau,p}^+;{\bf H}_{\tau,p}^-]$ (only has a single Fourier component) accordingly. 
Moreover, we have ${\bf H}_{\tau,p}^-=0$ if $p\leq {P}/2$ and ${\bf H}_{\tau,p}^+=0$ if $p>{P}/2$.
As indicated in Eq.~(\ref{Truextint}),
the approximate function of ${\bm e}_p={\mathcal{G}}_{\rm int}{\bm h}_p$ corresponds to a vector ${\bf e}_p={\bf T}_{\rm int}{\bf h}_p$,
where the vector ${\bf e}_p$ is a collection of the Fourier coefficients of tangential electric fields.
Therefore, the $p$-th column of the matrix ${\bf T}_{\rm int}$ is exactly ${\bf e}_p$ and we have
\begin{equation}\label{FormofTint}
	{\bf T}_{\rm int}=[{\bf e}_1,\cdots,{\bf e}_{{P}}].
\end{equation}

For given transverse magnetic fields ${\bm h}_p$,
the electric field ${\bf E}_p$ in $\Omega_D$ satisfying Eq.~(\ref{QPBC}) is the unique solution of the following boundary value problem
\begin{equation}\label{BVP}
	\begin{aligned}
		&\nabla\times\nabla\times{\bf E}_p-\left(\frac{\omega}{c}\right)^2\varepsilon(\br){\bf E}_p=0,\quad\br\in\Omega_D,\\
		&{\bm n}\times\frac{c}{{\rm i}\omega}\nabla\times{\bf E}_p={\bm h}_p\quad\text{ on }{\Gamma_\pm},
	\end{aligned}
\end{equation}
where $\bm n$ is the unit outer normal vector, i.e., ${\bm n}=[0;0;\pm 1]$ on $\Gamma_\pm$, respectively.
For a generic structure, we use a FEM to calculate the matrix ${\bf T}_{\rm int}$.
The solution ${\bf E}_p$ of problem (\ref{BVP}) is expanded by
\begin{equation}
	{\bf E}_p\approx\sum_{j=1}^{N}c_{pj}{\bm N}_j,
\end{equation}
where $N$ is the number of basis functions $\left\{{\bm N}_j\right\}$ and $c_{pj}$ are unknown coefficients. 
The coefficients satisfy the equation ${\bf K}{\bf c}_p={\rm i}\omega{\bf b}_p/c$,
where $\bf K$ is the sparse stiffness matrix and the right-hand side ${\bf b}_p$ is given by
\begin{equation}\label{rhs}
	\begin{aligned}
		&{b}_{pj}=-\int_{\Gamma_+} {\bf H}_{\tau,p}^+\cdot {\bm N}_{j}\,{\rm d}S\text{ if }p\leq {P}/2,\\
		&{b}_{pj}=-\int_{\Gamma_-} {\bf H}_{\tau,p}^-\cdot {\bm N}_{j}\,{\rm d}S\text{ if }p> {P}/2.
	\end{aligned}
\end{equation}
In fact, the vector ${\bf b}_p$ corresponds to a Fourier coefficient of basis functions which does not vanish on $\Gamma_\pm$.
The vector ${\bf e}_p$ is related to the following integrals
\begin{equation}\label{hlmn}
	{\bf E}_{\tau,p,mn}^\pm = \frac{1}{L^2}\sum_{j=1}^{N}c_{pj}\int_{\Gamma_\pm}{\bm N}_{j}\overline{F}_{mn}(x,y)\,{\rm d}S,
\end{equation}
where $\overline{F}_{mn}$ is the complex conjugate of $F_{mn}$.
According to Eqs.~(\ref{FormofTint}), (\ref{rhs}) and (\ref{hlmn}), we can show that
\begin{equation}\label{mrhsp}
	{\bf T}_{\rm int}=-\frac{{\rm i}\omega}{cL^2}{\bf B}^*{\bf K}^{-1}{\bf B},
\end{equation}
where ${\bf B}=[{\bf b}_1,\cdots,{\bf b}_{{P}}]$ and the symbol ``$*$'' represents the conjugate transpose of a matrix.
The matrix ${\bf B}$ is sparse since many basis functions vanish at the planes $\Gamma_\pm$.
It is clear that the matrix ${\bf T}_{\rm int}$ is the Schur complement of a sparse argument matrix $\hat{\bf K}$:
\begin{equation}
	\hat{\bf K}=\frac{{\rm i}\omega}{cL^2}\begin{bmatrix}
		{\bf K}& {\bf B}\\
		{\bf B}^*& {\bf O}
	\end{bmatrix},
\end{equation}  
where $\bf O$ is a zero matrix.
Thus, the matrix ${\bf T}_{\rm int}$ can be obtained by performing a partial factorization for the matrix $\hat{\bf K}$
\begin{equation}
	\hat{\bf K}=\frac{{\rm i}\omega}{cL^2}\begin{bmatrix}
		{\bf L}&{\bf O} \\
		{\bf L}_{21}& {\bf I}
	\end{bmatrix}\begin{bmatrix}
		{\bf U}&{\bf U}_{12} \\
		{\bf O}& {\bf T}_{\rm int}
	\end{bmatrix},
\end{equation}
where $\bf I$ is an identity matrix.
The factorization stops once the upper left block of $\hat{\bf K}$ is factorized into ${\bf K}={\bf L}{\bf U}$,
and can be handled by existing packages such as MUMPS~\cite{Amestoy2001} and PARDISO~\cite{Petra2014}.
Such a factorization avoids calculating the total coefficients ${\bf K}^{-1}{\bf B}$, and thus, the matrices $\bf L$ and $\bf U$
can be dropped during the factorization.
Existing numerical experiments show that calculating the Schur complement by the partial factorization procedure is more efficient with respect to CPU time and memory usage than the direct method based on computing ${\bf K}^{-1}{\bf B}$ by using LU factorization and forward and backward iterations~\cite{Petra2014,Hsu2022}.
\section{Contour integral method} 
Since the resonant modes are smooth on $\Gamma_\pm$, the integer $M$ in Eqs.~(\ref{Approximate}) and  (\ref{Truextint}) can be very small. 
Thus, the size of the matrix $\bf T$ is relatively small and the NEP (\ref{MatrixNEP}) can be solved efficiently by the contour integral method \cite{Asakura2009, Beyn2012}. 
In this section, we briefly review the contour integral method.
Suppose that the matrix ${\bf T}(\omega)$ is holomorphic in an open domain $\mathcal{D}$.
For a given smooth contour $\mathcal{C}\subset\mathcal{D}$, we assume that the matrix ${\bf T}(\omega)$ has no eigenvalues on the contour and only has semi-simple eigenvalues 
$\omega_k$, $k=1,\cdots,K$ (counted with multiplicity and $K\ll {P}$) in the interior $\mbox{int}(\mathcal{C})\subset\mathcal{D}$.
Moreover, we assume that the $K$ eigenvectors ${\bf h}_1,\cdots,{\bf h}_K$
are linearly independent, and the $K$ left eigenvectors are also linearly independent.
In applications, these assumptions are expected to hold in generic cases.

The steps of the contour integral method are listed as follows.
For an integer $Q>K$, we first chose a ${P}\times Q$ matrix ${\bf U}$ at random and compute the two integrals
\begin{equation}\label{A01}
	{\bf A}_0=\frac{1}{2\pi {\rm i}}\int_{\mathcal{C}}{\bf T}(\omega)^{-1}{\bf U}{\rm d}\omega,\quad
	{\bf A}_1=\frac{1}{2\pi {\rm i}}\int_{\mathcal{C}}\omega{\bf T}(\omega)^{-1}{\bf U}{\rm d}\omega.
\end{equation}
The matrix ${\bf A}_0$ has singular values
$
\sigma_1\geq\cdots\sigma_K>0=\sigma_{K+1}=\cdots=\sigma_{Q}.
$
Next, we compute the singular value decomposition of ${\bf A}_0$ in reduced form
$
{\bf A}_0={\bf V}_0{\bm \Sigma}_0{\bf W}_0^\dagger,
$
where ${\bm \Sigma}_0=\mbox{diag}(\sigma_1,\cdots,\sigma_{K})$, and build a $K\times K$ matrix ${\bf G}={\bf V}_0^\dagger{\bf A}_1{\bf W}_0{\bm \Sigma}_0^{-1}$.
As shown in Ref.~\cite{Beyn2012}, the eigenvalues of the matrix {\bf G} are the eigenvalues of the NEP (\ref{MatrixNEP}) inside the contour $\mathcal{C}$, and they share the same algebraic and geometric multiplicities.
Finally, we evaluate the eigenvalues $\omega_1,\cdots,\omega_{K}$ and their corresponding eigenvectors ${\bf g}_1,\cdots,{\bf g}_K$ of the matrix $\bf G$.
The eigenvector ${\bf h}_k$ of the NEP (\ref{MatrixNEP}) can be obtained by ${\bf h}_k={\bf V}_0{\bf g}_k$.

The two integrals in Eq.~(\ref{A01}) can be approximated by numerical quadrature,
and more specifically, if the contour $\mathcal{C}$ is a circle with a parameterization:
\begin{equation}\label{circlepara}
	\omega(\theta)=\omega_0+\rho e^{{\rm i}\theta},\quad 0\leq\theta<2\pi,
\end{equation}
we have
\begin{equation}
	{\bf A}_{0,N_q}=\frac{\rho}{N_q}\sum_{q=0}^{N_q-1}{\bf T}[\omega(\theta_q)]^{-1}{\bf U}e^{{\rm i}\theta_q},\quad 
	{\bf A}_{1,N_q}=\frac{\rho^2}{N_q}\sum_{q=0}^{N_q-1}{\bf T}[\omega(\theta_q)]^{-1}{\bf U}e^{2{\rm i}\theta_q}+\omega_0{\bf A}_{0,N_q},
\end{equation}
where $\theta_q=(2q+1)\pi/N_q$.
Once the eigenvalue $\omega_k$ and eigenvector ${\bf h}_k$ are obtained, the electric field ${\bf E}_k$ in $\Omega_D$ can be obtained by
\[
{\bf E}_k = \sum_{j=1}^Nc_{kj}{\bm N}_j,\quad {\bf c}_k={\rm i}\frac{\omega_k}{c}{\bf K}(\omega_k)^{-1}{\bf B}{\bf h}_k.
\]
The electric fields in $\Omega_\pm$ can be obtained from the Rayleigh expansion using definition of exterior TI operators.
\section{Numerical Examples}
\begin{figure}[h]
	\centering
	\includegraphics[scale=1.2]{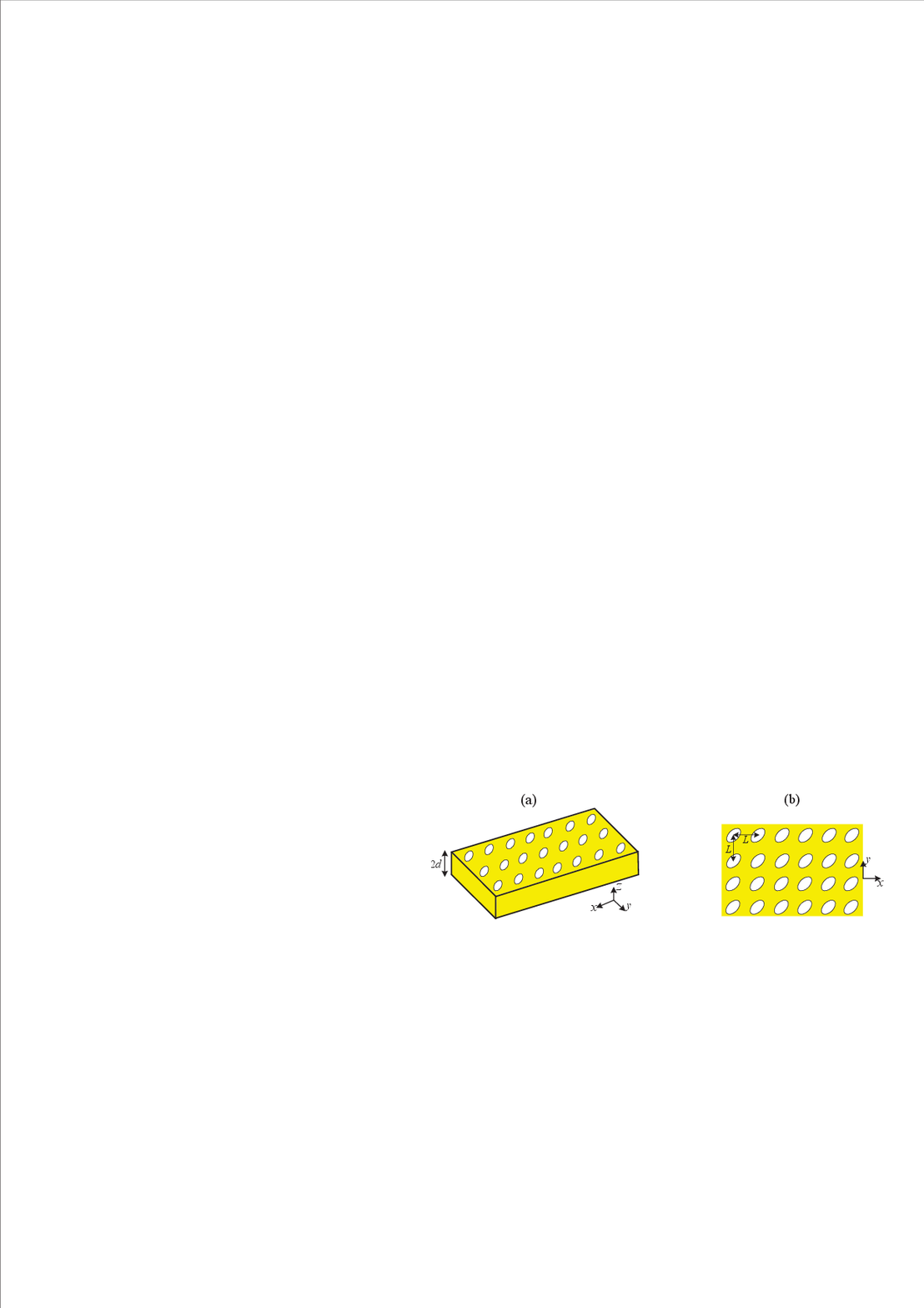}
	\caption{A PhC slab with a square lattice of elliptic air holes: (a) 3D
		view and (b) top view.}\label{PhCSlab}
\end{figure}
In order to verify the accuracy and efficiency of our method, 
we calculate resonant modes in a photonic crystal (PhC) slab and a square lattice of dielectric spheres on top of a uniform slab.
We first consider resonant modes in a PhC slab. 
As shown in Fig.~\ref{PhCSlab}, 
the PhC slab is a uniform slab (surrounded by air) with a square lattice of elliptic air holes. 
The thickness of the PhC slab is $2d=0.5L$.
The lengths of semimajor and semiminor axes of the elliptic air holes are $0.3L$ and $0.2L$, respectively, and the angle between the major axis
and the $x$ axis is $\pi/6$.
The dielectric constant of the slab is 4.0804 (silicon nitride).
The structure is symmetric in $z$ and we consider resonant modes with $\alpha L/2\pi=-0.13$, $\beta L/2\pi=0.18$ and assume $\mathcal{P}{\bf E}={\bf E}$.
The size of the TI matrices can be reduced to ${P}/2\times {P}/2$ by taking advantage of symmetry in $z$.

We chose $D=0.3L$, and adopt the first kind lowest-order N\'{e}d\'{e}lec element to compute the interior TI matrix ${\bf T}_{\rm int}$.
We use the contour integral method to calculate resonant modes inside a circle $\mathcal{C}$ with center at $\omega_0 L/2\pi c=0.684$ and radius $\rho L/2\pi =8\times 10^{-3}$. 
The number of the quadrature nodes is $N_q=21$.
Numerical results show that there exists only one resonant mode inside the circle.
We emphasize that the method does not produce spurious solutions.
Tables \ref{Nx1real} and \ref{Nx1imag} 
\begin{table}[h]
	\newcommand{\tabincell}[2]{\begin{tabular}{@{}#1@{}}#2\end{tabular}}
	\centering 
	\caption{The real part of the normalized frequency $\omega L/2\pi c$($\times 10^{-1}$).}
	\begin{tabular}{|c|c|c|c|c|c|c|c|}
		\hline
		$h/L$& $M=6$ & $M=5$ & $M=4$ & $M=3$ & $M=2$ &$M=1$ \\
		\hline
		$5.00\times 10^{-2}$  &6.9167& 6.9167& 6.9167& 6.9167& 6.9168& 6.9153\\
		$2.50\times 10^{-2}$  &6.9220& 6.9220& 6.9220& 6.9220& 6.9220& 6.9205\\
		$1.25\times 10^{-2}$  &6.9234& 6.9234& 6.9234& 6.9234& 6.9234& 6.9218\\
		$6.25\times 10^{-3}$  &6.9237& 6.9237& 6.9237& 6.9237& 6.9238& 6.9222\\
		\hline                                                              
	\end{tabular}\label{Nx1real}                                    
\end{table}
\begin{table}[h]
	\newcommand{\tabincell}[2]{\begin{tabular}{@{}#1@{}}#2\end{tabular}}
	\centering 
	\caption{The imaginary part of the normalized frequency $\omega L/2\pi c$ ($\times 10^{-4}$).}
	\begin{tabular}{|c|c|c|c|c|c|c|c|}
		\hline
		$h/L$& $M=6$ & $M=5$ & $M=4$ & $M=3$ & $M=2$ &$M=1$ \\
		\hline
		$5.00\times 10^{-2}$  &3.670& 3.670& 3.670& 3.669& 3.667& 3.651\\
		$2.50\times 10^{-2}$  &3.569& 3.569& 3.569& 3.568& 3.565& 3.551\\
		$1.25\times 10^{-2}$  &3.544& 3.544& 3.544& 3.544& 3.541& 3.526\\
		$6.25\times 10^{-3}$  &3.539& 3.539& 3.539& 3.538& 3.535& 3.521\\
		\hline
	\end{tabular}\label{Nx1imag}
\end{table}
show the real and imaginary parts of the normalized frequency $\omega L/2\pi c$ of the resonant mode for different $M$ and mesh size $h$.
It can be seen that the convergence in $M$ is much faster than that in the mesh size $h$.
Therefore, we only need a relatively small TI matrix to retain the accuracy even though $h$ can be very small.
Moreover, for a fixed mesh size $h$, we have chosen a large value $N_q=21$ to solve the NEP (\ref{MatrixNEP}) with very high accuracy (twelve digits).
In fact, considering the error of domain discretization, a smaller $N_q$ is enough to obtain a result with relatively high accuracy.
For example, for $M=6$ and $h/L=2.5\times 10^{-2}$, we obtain $\omega L/2\pi c=6.9220\times 10^{-1}-{\rm i}3.569\times 10^{-4}$ with $N_q=11$, same as the result for $N_q=21$.

To illustrate the efficiency of our method, we also calculate the resonant mode using a similar FEM and PMLs.
The unbounded domain is truncated to a bounded domain $\Omega_{z_*}=S\times(0,z_*)$.
The interval $(0,z_*)$ is divided into three sub-intervals, i.e., $(0,z_1)$, $(z_1, z_2)$ and $(z_2,z_*)$, where
$z_1=d$, $z_2=d+1.5L$, and $z_*=d+2L$.
The intervals $(z_1, z_2)$ and $(z_2,z_*)$ correspond to air layer and the PML, respectively.
In the PML, the variable $z$ is transformed to a stretching complex variable $\hat{z}$ given by
\[
\hat{z}=z_2+\int_{z_2}^{z}s(\tilde{z})d\tilde{z},\quad s(\tilde{z}) = 1 + {\rm i}183\left(\frac{\tilde{z} - z_2}{z_3-z_2}\right)^3.
\]
The resulting linear generalized matrix eigenvalue problem is solved by the Arnoldi method.
For $h/L=2.5\times 10^{-2}$, we obtain $\omega L/2\pi c=6.9225\times 10^{-1} - {\rm i}3.612\times 10^{-3}$.
It can be seen that the results calculated by the two methods are consist.

For the same mesh size, the number of unknowns $N$ in the method with PMLs is much larger than the method without PMLs.
For example, when $h/L=2.5\times 10^{-2}$, we have $N=908,061$ for the method with PMLs but $N=230,121$ for the method without PMLs.
In this case, for the method with PMLs, the CPU time is 44 min and the peak memory required is 64 GB.
However, for the method without PMLs, the CPU time is 7 min the peak memory required is 12 GB with $N_q=21$ and $M=6$.
Moreover, the CPU time is only 4 min with $N_q=11$.
It is clear that the method without PMLs is indeed efficient with respect to CPU time and memory usage compared with the method using PMLs.
The computational platform is provided by the centralised high performance computing (HPC) at City University of Hong Kong.
All the numerical experiments are run in serial on an exclusive node with specifications: two AMD EPYC 7522 processors with a total of 128 cores.

\begin{figure}[h]
	\centering
	\includegraphics[scale=0.6]{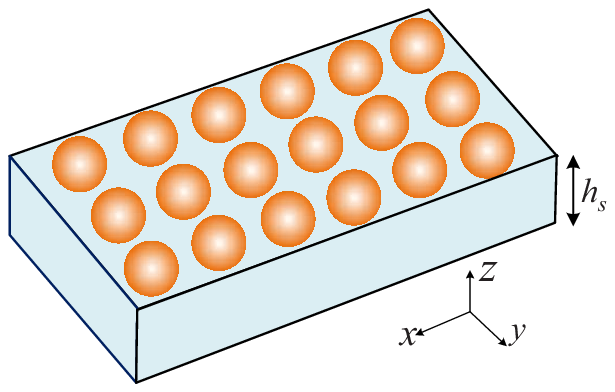}
	\caption{A square lattice of dielectric spheres on top of a uniform dielectric slab.}\label{SlabSphere}
\end{figure}

\begin{figure}[h]
	\centering
	\includegraphics[scale=0.32]{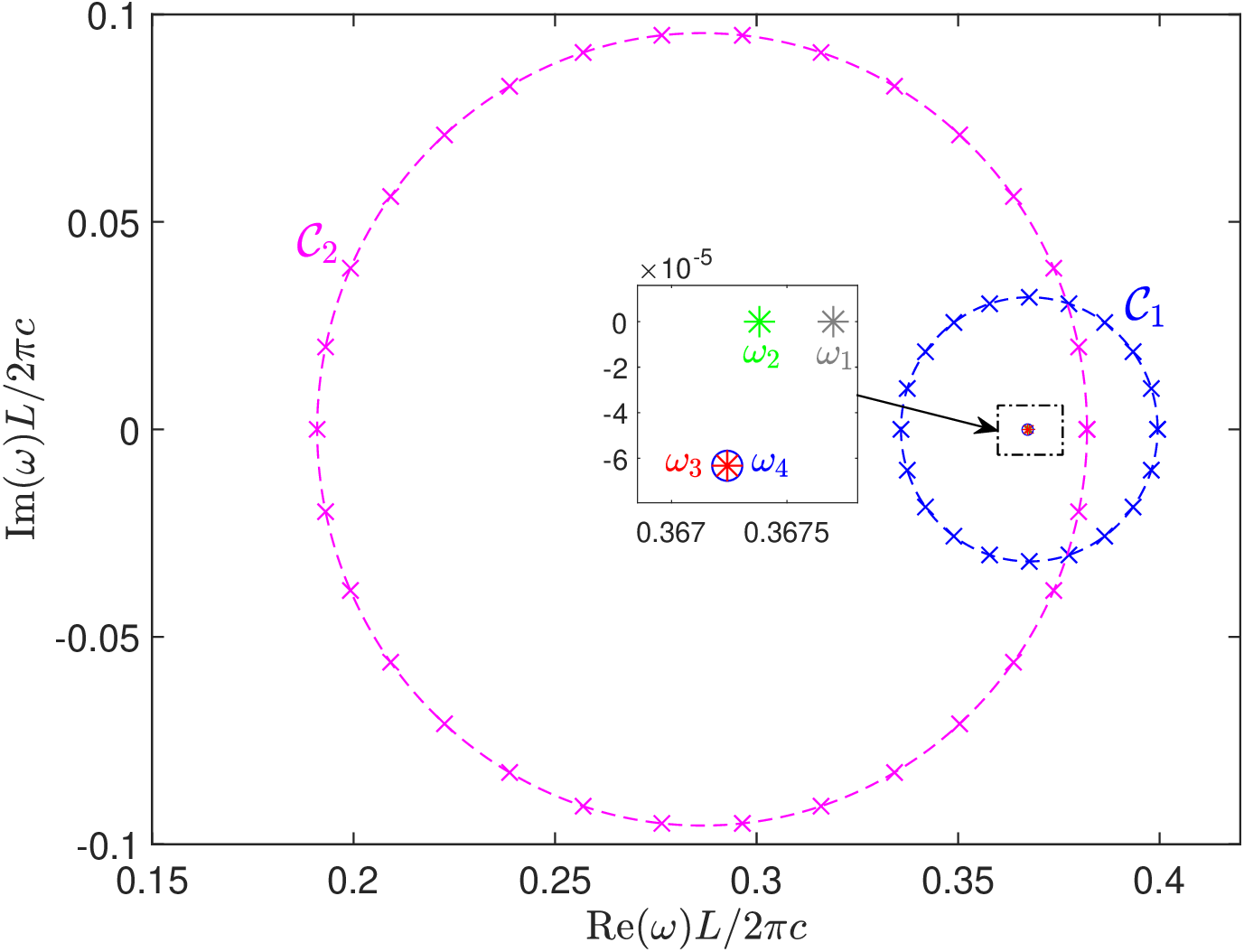}
	\caption{Complex $\omega$ plane with integration
		contours $\mathcal{C}_1$ (blue) and $\mathcal{C}_2$ (pink). 	
		The frequencies $\omega_k$, $k=1,2,3$ (asterisks) and $\omega_4$ (circle) calculated using contour $\mathcal{C}_2$ are shown in the inset. 
		The results obtained by contours $\mathcal{C}_1$ and $\mathcal{C}_2$ are nearly identical.}\label{Contour}
\end{figure}

\begin{figure}[h]
	\centering
	\includegraphics[scale=0.35]{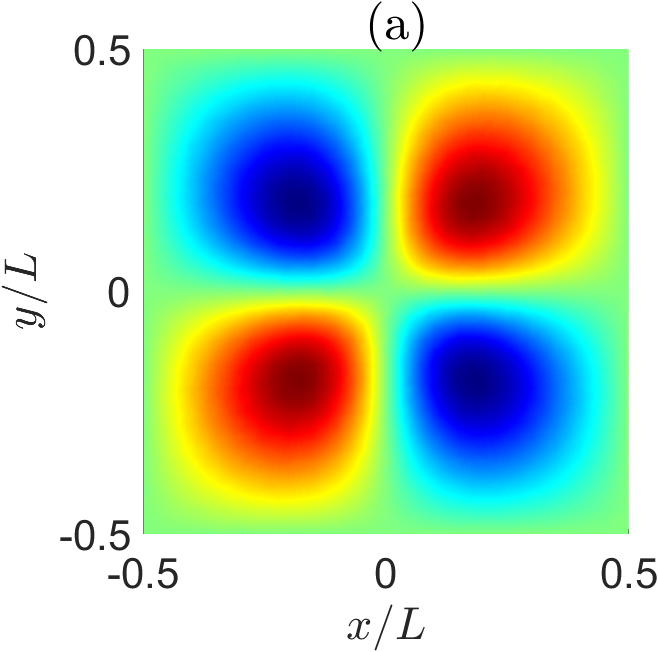}
	\includegraphics[scale=0.35]{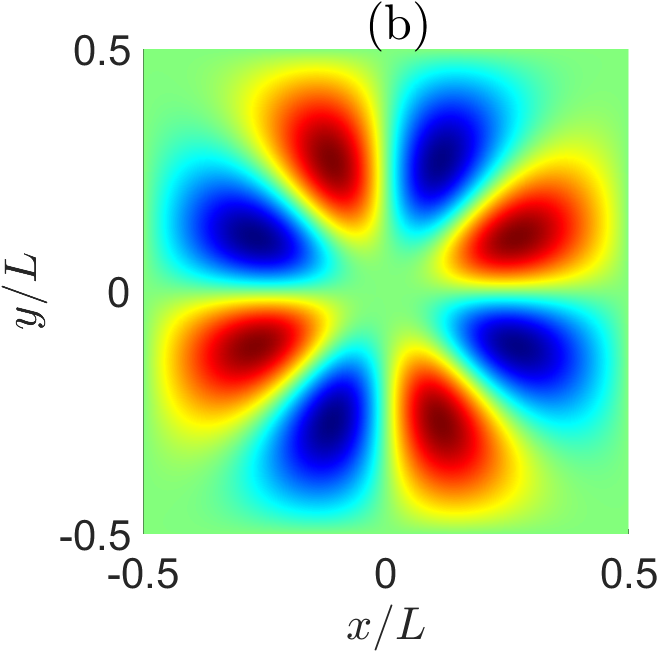}
	\includegraphics[scale=0.35]{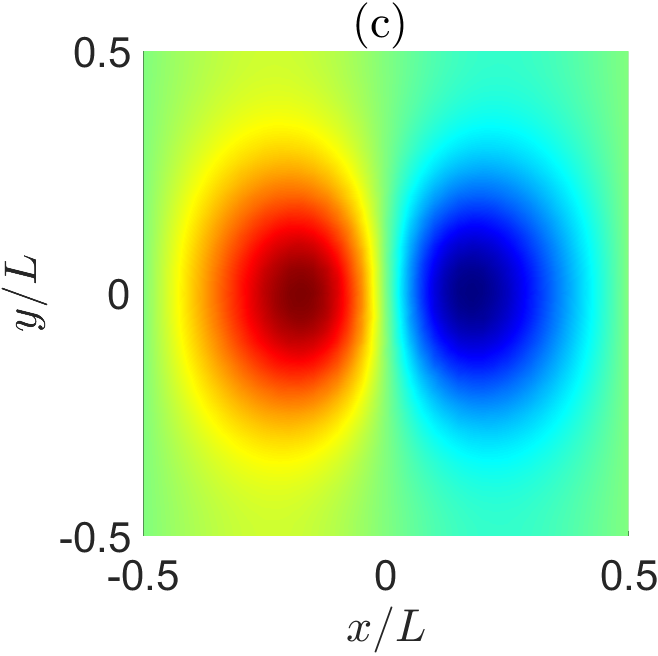}
	\includegraphics[scale=0.35]{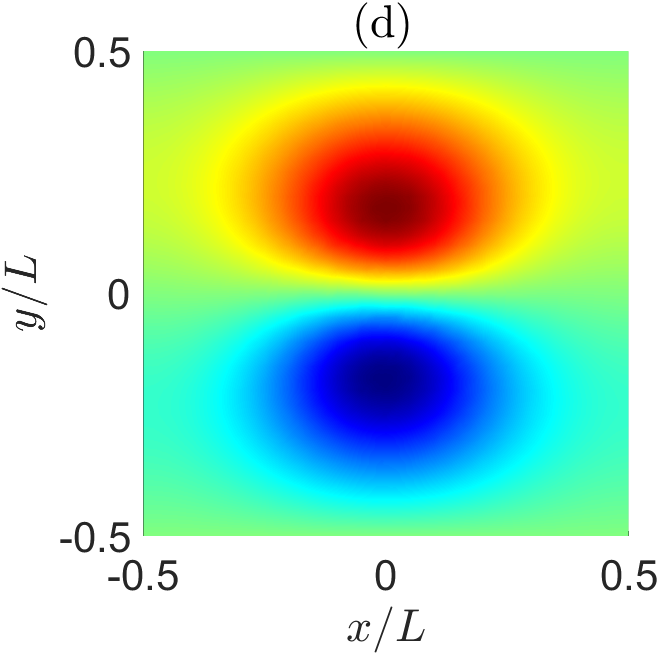}
	\caption{$\mbox{Im}(E_z)$ of four resonant modes at $z=D$.}\label{RMfield}
\end{figure}

Next, we consider resonant modes with $\alpha=\beta=0$ in a square lattice of dielectric spheres on top of a uniform slab shown in Fig.~\ref{SlabSphere}.
The radius of spheres is $0.3L$.
The thickness of the slab is $h_s=0.3L$.
The dielectric constants of the slab and spheres are 12.96 (silicon) and 4.0804 (silicon nitride), respectively.
Let $d=0.45L$ and $D=0.5L$.
We adopt the first kind quadratic N\'{e}d\'{e}lec element to approximate the interior TI matrix ${\bf T}_{\rm int}$.
We use the contour integral method to calculate the resonant modes inside two circles $\mathcal{C}_1$ and $\mathcal{C}_2$ with centers at $\omega_0 L/2\pi c=0.368$, $0.286$ and radii $\rho L/2\pi =3.18\times 10^{-2}$, $9.55\times 10^{-2}$, respectively. 
The numbers of the quadrature nodes are $N_q=21$ and $31$, respectively.
As shown in Fig.~\ref{Contour}, for $h/L=5\times 10^{-2}$, we only obtain four resonant modes inside these two circles and our method does not produce spurious solutions.
Moreover, the results are not affected by the location of the contour.
Namely, our method is robust with respect to the location of the contour (if it still contains the resonant modes).
Table~\ref{Nx2} shows the normalized frequency $\omega L/2\pi c$ of four resonant modes for different mesh sizes $h$.
The imaginary part of $E_z$ at plane $z=D$ for four resonant modes is shown in Fig.~\ref{RMfield}.
\begin{table}[h]
	\newcommand{\tabincell}[2]{\begin{tabular}{@{}#1@{}}#2\end{tabular}}
	\centering 
	\caption{Normalized frequencies $\omega L/2\pi c$ of four resonant modes with different mesh size $h$.}
	\begin{tabular}{|c|c|c|}
		\hline
		$h/L$ &  $5.0\times 10^{-2}$ &  $2.5\times 10^{-2}$ \\
		\hline
		$\omega_1 L/2\pi c$&\hspace{0.4mm} $3.67701\times 10^{-1} - {\rm i}7.91\times 10^{-14}$&\hspace{0.4mm} $3.67697\times 10^{-1} - {\rm i}1.71\times 10^{-14}$\\
		\hline
		$\omega_2 L/2\pi c$&\hspace{0.4mm} $3.67381\times 10^{-1} - {\rm i}9.78\times 10^{-13}$&\hspace{0.4mm} $3.67375\times 10^{-1} - {\rm i}1.36\times 10^{-13}$\\
		\hline
		$\omega_3 L/2\pi c$&$3.67241\times 10^{-1}- {\rm i}6.33\times 10^{-5}$&$3.67235\times 10^{-1} - {\rm i}6.38\times 10^{-5}$\\
		\hline
		$\omega_4 L/2\pi c$&$3.67241\times 10^{-1}- {\rm i}6.33\times 10^{-5}$&$3.67235\times 10^{-1} - {\rm i}6.38\times 10^{-5}$\\
		\hline                                                              
	\end{tabular}\label{Nx2}                                    
\end{table}
Our results are consistent with the results reported in Ref.~\cite{Zhang2015}. 

Due to the very small imaginary part of $\omega_1$ and $\omega_2$, the first and the second resonant modes are actually BICs~\cite{bonnet1994,Hsu16NRM,Shipman2003}.
A BIC can be regarded as a special resonant mode with $\mbox{Im}(\omega)=0$.
The third and fourth resonant modes are doubly-degenerate resonant modes since they have different mode patterns and almost identical resonant frequencies.
Our method is efficient to determine degenerate resonant modes since the contour integral method inherits the algebra and geometric properties of the eigenvalues of the NEP (\ref{MatrixNEP}).
\section{Conclusion}
In this paper, we developed a new method without PMLs to calculate resonant modes in biperiodic structures.
We divide a unit cell into two exterior and one interior subdomains, calculate exterior and interior TI matrices, convert the original eigenvalue problem to a matrix NEP, and solve the NEP by the contour integral method.
Since the resonant modes are smooth at the interfaces between subdomains, the TI matrices can be relatively small.
The exterior TI matrix is obtained directly since the fields have explicit expressions in the exterior subdomains.
For a generic structure, we adopted a fast and memory-efficient method to calculate the interior TI matrix.
Compared with a FEM using PMLs, numerical examples show that our method has advantages with respect to memory usage and CPU time.
Moreover, our method is free of spurious solutions, and is also efficient to determine degenerate resonant modes.
We believe that our new method holds great potential for real-world applications.
\bibliography{refs}
\end{document}